\documentclass[preprint,twocolumn]{osajnl}
\journal{ol} 
\setboolean{shortarticle}{true} 

\title{ Nonlinear imaging through a golden spiral multicore fiber}
\author[1]{Siddharth Sivankutty}
\author[1]{Viktor Tsvirkun}
\author[2] {Olivier Vanvincq}
\author[2]{G\'{e}raud Bouwmans}
\author[2,3]{Esben Ravn Andresen}
\author[1,*]{Herv\'{e} Rigneault}
\affil[1]{Aix Marseille Univ, CNRS, Centrale Marseille, Institut Fresnel, F-13013 Marseille, France}
\affil[2]{Univ. Lille,CNRS, UMR 8523, Laboratoire de Physique des Lasers Atomes et Molécules, F-59000 Lille, France}
\affil[3]{Email: esben.andresen@univ-lille1.fr}
\affil[*]{Corresponding author: herve.rigneault@fresnel.fr}
\dates{Compiled \today}
\ociscodes{060.2430, 060.2350, 170.2150, 180.5810, 080.1238, 110.1080}

\begin{abstract}
We report two-photon lensless imaging through a novel golden spiral multicore fiber. This unique layout optimizes the sidelobe levels, field of view, cross-talk, group delay and mode density to achieve a sidelobe contrast of atleast 10.9 dB. We demonstrate experimentally the ability to generate and scan a focal point with a femtosecond pulse and perform two-photon imaging.
\end{abstract}
\begin{document}

\maketitle

Fiber bundles are key tools in the realization of miniaturized imaging systems for imaging in hard-to-access regions, such as \textit{in vivo} endoscopy. In this context, there has been a keen interest in extreme miniaturization of the imaging systems with no opto-mechanical elements at the distal end. This ensures  fiber-based probes are minimally invasive, opening the route for imaging regions hitherto considered impossible, such as imaging deep and interconnected regions of the brain. While fiber bundle systems that act like traditional cameras are already prevalent, the adoption of wavefront shaping techniques with bare fibers have emerged as a promising route. This class of imaging techniques called lensless endoscopy, offer additional advantages: pixelation-free imaging, diffraction-limited resolution and fields of view that limited only by the fiber numerical aperture (NA) \cite{andresen2016ultrathin,Thompson2011,Andresen2013a,CizmarNatComm2012,conkey2016lensless}. Hence the ability to control the wavefront at the tip of the fiber, either with multicore or multimode fibers offer new features such as axial sectioning and the hope for chemical specificity and label-free imaging with nonlinear contrast mechanisms. 

Multicore fibers (MCF) with single mode cores  are promising candidates for the realization of lensless endoscopes working with nonlinear contrast mechanisms. This primarily stems from its ability to deliver ultrashort laser pulses and places it apart from multimode fibers whose modal dispersion quickly becomes deleterious.  In addition, multicore fibers with widely spaced single mode cores exhibit very weak coupling and have proven to be robust to bending in terms of spatial \cite{Kim2015}, temporal \cite{tsvirkun2017bending} and polarization distortion \cite{Sivankutty:16}. For a comprehensive review of MCFs in nonlinear imaging, we refer the interested readers to \cite{andresen2016ultrathin}.

The performance of the lensless endoscope is highly dependent on the layout of the single mode cores. For a MCF with identical cores, we can write the farfield intensity pattern as the product of the diffracted intensity pattern of a single core and an interference factor $\mathrm{IF}$\cite{Sivankutty2016OLprbundle}.
 
\begin{align}
  \mathrm{I(k)} &=   \mathrm{I_{\mathrm{0}}(k)} \mathrm{IF(k)}    
    \label{eqn:FF}
\end{align}
This interference term is related to the AF, the geometric layout of the fiber cores, by the Wiener-Khinchin theorem as  
\begin{align}
\mathrm{IF} &= \left|\mathcal{F}\Big[AF\Big]\right|^2
  \label{eqn:WK}
  \end{align}

\begin{figure}[t]
  \centering
  \includegraphics[width =0.9\columnwidth]{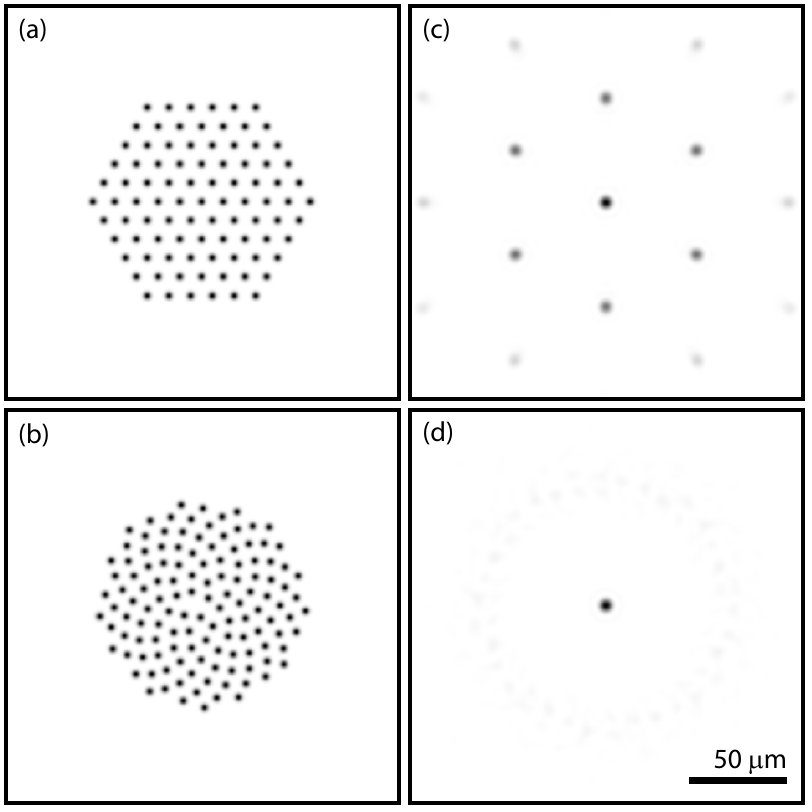}
  \caption{
Calculations: (a) Layout of the fiber cores in a hexagonal tiling and (b) in a Fermat's spiral. (c)-(d)Simulated intensity PSFs with an intermediate focus at a distance z = $500~\mu$m from the fiber for  the hexagonal and the Fermat's spiral arrangement.
\label{fig:Sims}}
\end{figure}

\begin{figure}[t]
  \centering
  \includegraphics[width =0.9\columnwidth]{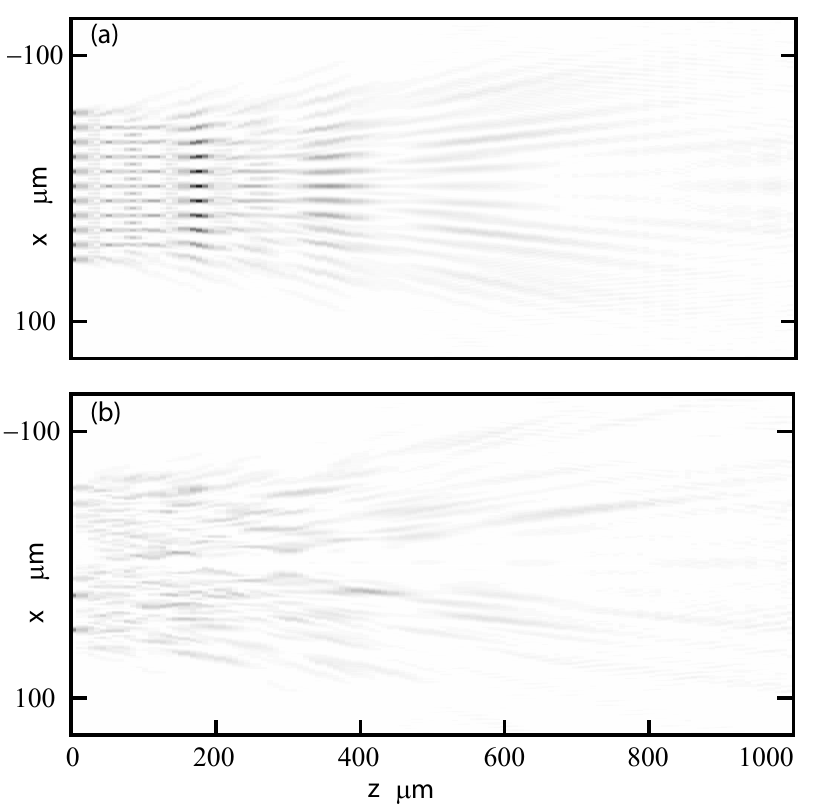}
  \caption{
Calculations: (a) An xz cut of the diffracted intensity pattern by the arrangement of cores in a hexagonal arrangement where intense focii appear periodically in x- and z- directions (b) The same plot for the cores arranged in a Fermat's spiral where the focii are replaced by a broader speckle background. Note both images are plotted on the same color scale.
\label{fig:Sim2}}
\end{figure}

From Equation \ref{eqn:WK} it is clear that any periodicity in the layout of the cores will manifest as sharp peaks in  $\mathrm{IF}$. For example, if the positions of the cores are described by a comb function, $\operatorname{III}_\Lambda\left(x\right)$, where $\Lambda$ is the pitch, the resulting diffraction pattern will exhibit spatial frequency components at $\frac{2\pi n}{\Lambda}$. This results in replicas of the object when the focus is being scanned across it. Hence, we seek a singly-peaked $\mathrm{IF}$, which corresponds to a highly disordered arrangement of the cores. In \cite{Sivankutty2016OLprbundle} we introduced a method to realize pseudo-random positioning of the cores by introducing a deterministic radial offset and a random angular rotation around a master lattice and was further extended to rotation of groups of fibers in \citep{kim2018semi}. The scientific literature on optimizing the array layouts is rich and includes several strategies starting from purely optimization based approaches, stochastic models and elegant analytical solutions for aperiodic tiling \cite{mailloux2005phased}. Here, we report a 120-core MCF designed with a deterministic aperiodic tiling based on the Fermat's spiral. The positions of the cores in a Fermat spiral follow the relations:
\begin{align}
  \rho_n &= \Lambda \sqrt{n},\,\,\,\,\,\,\,\,
  \theta_n =   n\pi (3-\sqrt{5}),
\label{Eqn:Fermat}
\end{align}
where $\rho_n$ is the radial distance of the $\mathit{n^{th}}$ element and $\theta_n$ is its corresponding angular coordinate. The parameter $\Lambda$ is a free parameter that controls the minimal spacing between the cores and the golden angle constant, $(3-\sqrt{5})$, results in non-redundant positions of the cores. In addition, the Fermat spiral array layout also has  a high packing fraction~\cite{vogel1979better}. We note that this design was recently proposed by Gabrielli \textit{et  al.} in the context of nano-antenna arrays \cite{gabrielli2016aperiodic} and demonstrated on an 8-element array \cite{pita2017side}.
A key difference in our work, unlike in typical phased array configurations, the lensless endoscope operates in the Fresnel regime with only an imperfect overlap of the individual beamlets from the cores. Hence, we simulate the performance of the golden spiral array and the conventional hexagonal array with the angular spectrum approach in this intermediate regime. Figure \ref{fig:Sims}(a-b) shows the geometry of the considered MCF in both the configurations with a divergence of 0.12 rad, and Figures \ref{fig:Sims}(c-d) are images of the  diffraction patterns at a distance $500 ~\mu$m away from the MCF facet. Note that an additional quadratic phase is imposed to generate a focal spot at this plane to enable a visual comparison of the eventual point spread function (PSF) of the lensless endoscope in either case.

Since the NA of the individual cores and the minimal distance between them ($\Lambda$) are chosen to support only the fundamental mode of the core and ensure no inter-core coupling~\cite{snyder1972coupled}, all the parameters of the spiral naturally emerge with no further need to optimize any design parameters. Clearly, given a degree of overlap of the individual beamlets, the golden spiral MCF exhibits a PSF that is centrally peaked [Figure~\ref{fig:Sims}(d)] in comparison with the PSF with multiple satellite peaks associated with the hexagonal array [Figure~\ref{fig:Sims}(c)]. We also highlight in the case of the hexagonal array, the axial diffracted intensity patterns depicted in  Figure~\ref{fig:Sim2}(a) exhibits intense on-axis and satellite focii. In comparison, the intensity pattern of the Fermat spiral exhibits no such self imaging of the array itself and no significant on/off-axis focii appear in Figure~\ref{fig:Sim2}(b). This is the key insight: the aperiodicity offered by the Fermat spiral ensures there are no common positions in space where all the beamlets are in phase due to the geometric reasons (See Visualization 1). This is clear in Figure~\ref{fig:Sims}(c-d) where the energy of the incident beamlets are localized to six intense satellite focal spots occupying an effective area of $48 ~\mu\mathrm{m^2}$  whereas it is distributed over an effective area of $240~\mu \mathrm{m^2}$ for the Fermat spiral. Hence, given our experimental constraints, we identify the Fermat spiral as a near optimal design for sidelobe reduction in lensless endoscopes whilst maintaining a high packing efficiency.

\begin{figure}[t]
	\centering
	\includegraphics[width = 1\columnwidth]{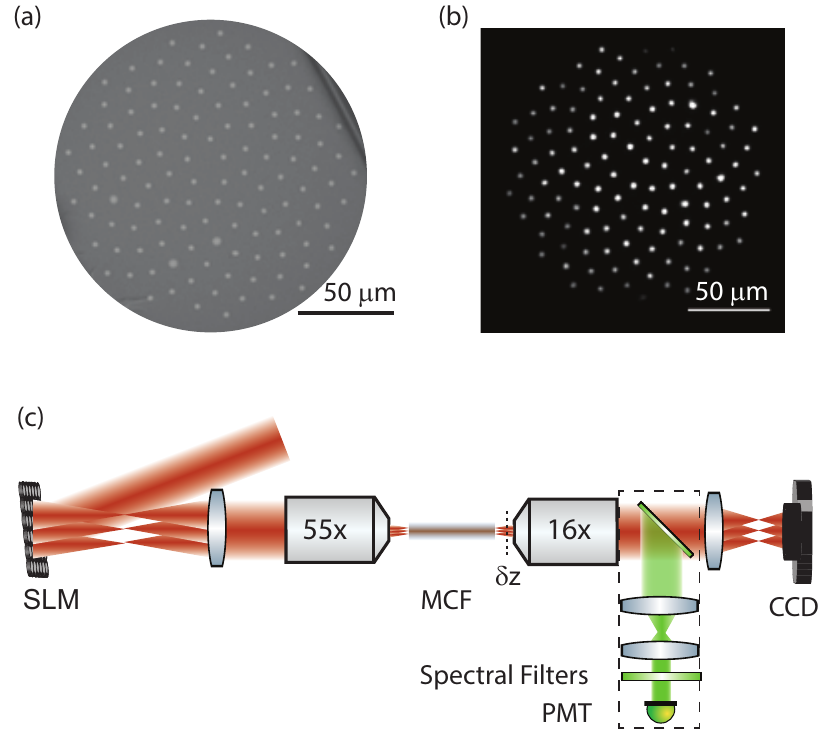}
  \caption{
(a) An electron micrograph of the golden spiral MCF. (b) End facet of the fiber with laser light coupled. (c) A simplified view of the setup used to characterize the imaging performance of the MCF.
\label{fig:Setup}}
\end{figure}

We fabricated a novel 120-core golden spiral MCF as depicted in the electron micrograph in Figure~\ref{fig:Setup}(a) in the following fashion: initially 120 holes of 2.1 mm diameter following the Fermat spiral are drilled into a 50 mm silica rod upto a depth of 230 mm. The standard deviations of the positions from the Fermat spiral on both the inlet and outlet face with respect to the drilling are only around $4{\times}10^{-3}$~mm and $9{\times}10^{-2}$~mm respectively while the standard deviation on the diameter is about $15 {\times} 10^{-3}$~mm. Germanium doped glass rods drawn from a preform  with a maximum index difference of $30{\times}10^{-3}$ w.r.t. silica (parabolic index - Prysmian Group) are then inserted into these holes. This preform is drawn in two steps into a MCF with an outer diameter = 176 $\mu$m, fiber core diameter core = $3.6~\mu$m and a beam divergence of 0.11 rad and the average nearest-neighbour distance, $\Lambda$~is determined to be $11.8~\mu$m 
We measure the average inter-core group delay dispersion in a 400 mm long MCF to be less than 128 fs which is less than the initial pulse width. This is a result of the very high homogeneity of the group indices of the cores. The measured group delay distribution is comparable to the conventional MCFs in a hexagonal tiling \cite{Andresen2015a}. This indicates that there are no additional stress-related distortions during the fabrication of the drilled preform or in the drawing process in these golden spiral MCFs. 

In order to characterize the sidelobe levels and the imaging performance with pulsed light, we work with an experimental setup whose simplified schematic is presented in Figure~\ref{fig:Setup}(c). A key element of the setup is a liquid crystal spatial light modulator (SLM, X10468, Hammamatsu) upon which the phase of a lenslet array is  inscribed. The positions of the individual lenses of this array match the positions of the fiber cores themselves. The source is a pulsed Ti:Sapphire laser (Chameleon, Coherent Inc.) operating at $920$ nm, delivering 150 fs pulses at 80 MHz. After reflection off the SLM, the laser beam converges to an array of spots arranged in the Fermat spiral. A futher demagnification by a factor 55 matches the focii array  to the size and the NA of the fiber cores themselves. In this study we used a 400 mm long fiber held relatively straight to preclude any temporal distortion of the pulses. The fiber end facet is imaged onto a CMOS camera (Flea3, FLIR) with a magnification of 16.5x. The MCF distal end is mounted on a translation stage such that we can image axial planes a few hundred micrometers away from the facet as well to examine their diffraction patterns.
  
\begin{figure}[t]
  \centering
  \includegraphics[scale = 1]{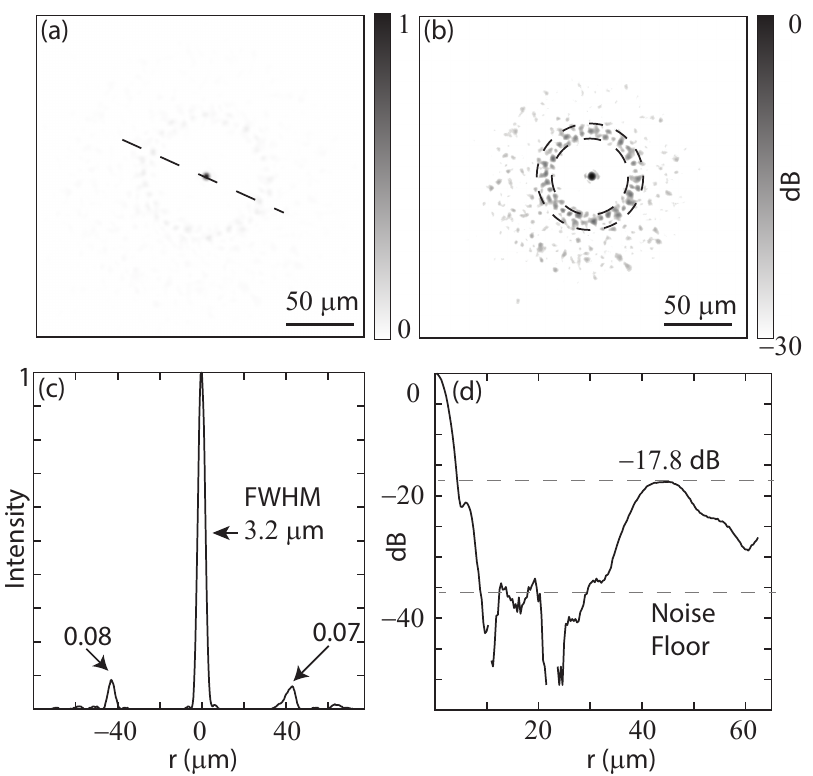}
  \caption{
(a) Experimental intensity images at a plane $500 ~\mu$m away from the fiber endface. (b) The corresponding log-plot highlighting the sidelobes at $\approx 45 ~\mu$m and the nonspecific speckle background.(c) An intensity plot of the central and the highest sidelobes indicated by the dashed line in panel~(a). (d)~The azimuthal average of the intensities of the speckle grains denoted by the annulus in panel~(b).
\label{fig:Expt2D}}
\end{figure}

At 920 nm, we observe the  $\mathrm{LP_{11}}$ mode group when the launch spot is offset from the center of the individual fiber cores. Hence, we take care to center the focal spots to the fiber cores to avoid exciting these modes. Figure~\ref{fig:Setup}(b) is an example when the coupling is optimal and no higher order modes are visible. This ensures we can phase all the fundamental $\mathrm{LP_{01}}$ modes with very high fidelity employing a simple piston term on each virtual lenslet using the SLM. After traversing the bundle, light in each core acquires a random phase difference w.r.t. to a reference (typically the central core). This manifests as an intensity distribution resembling a speckle at a plane $500 ~\mu$m away from the center. So an initial compensation of this phase distortion is performed as reported in \citep{Andresen2013a}. Since the SLM only works with \textit{p-}polarized light and the output polarization state of each of the beamlets is random \cite{Sivankutty:16}, we choose to work only with a single component of polarization. This results in the intensity patterns visualized in the Figure~\ref{fig:Expt2D}(a-b) at a plane $500 ~\mu$m away from the fiber. We choose this plane in particular, since it offers the optimal combination in terms of overlap of the individual beamlets from the cores thereby maximizing the Strehl ratio, the resolution and FoV for the imaging experiments. We recall that the resolution and FoV parameters are inversely proportional to one another. Moreover any effort to increase  the FoV by simply moving farther from the MCF endfacet would adversely affect the Strehl ratio and consequently the SNR of the two-photon imaging~\cite{Sivankutty2016OLprbundle}.

The main metric of interest for us is the sidelobe levels in the Fresnel regime. Figure~\ref{fig:Expt2D}(a) is an example of such a generated focus. In addtion to the intense central peak, we also observe the characteristic array factor of the golden spiral, with the sidelobes appearing at a radius $\approx 45 ~\mu$m that appear in the log plot in Figure ~\ref{fig:Expt2D}(b). The highest of the sidelobes exhibit a contrast of -10.9 dB compared to the peak [Figure ~\ref{fig:Expt2D}(c)] and the average of the highest thirty speckle grains in the first ring is measured to be  approximately -12 dB. In comparison, for the true far-field operation, we measure the peak sidelobe to be at -11.7 dB (data not shown). We observe that there is a marginal increase of the sidelobe levels and the nonspecific background speckle as compared to the far-field operation. We interpret this as the result of the imperfect spatial overlap of the individual beamlets in this intermediate regime. Another clear advantage of the Fermat spiral shows up as the highly contrasted region between the central lobe and the sidelobes even in this intermediate regime. In the earlier approaches \cite{Sivankutty2016OLprbundle, kim2018semi}, such a region could not be engineered due to inherent randomness of the design of the array.

From Figure \ref{fig:Expt2D}(c), the FWHM of the central lobe is measured to be $3.2 ~\mu$m, and in turn the resolution of our eventual two-photon images to be $\approx ~2.2 ~\mu$m. When we calculate the azimuthal average of the intensity over the annulus corresponding to the smeared sidelobes, we obtain an average sidelobe level of -17.8 dB [Figure \ref{fig:Expt2D}(d)]. Since the nonlinear signal is dependent on the peak irradiation, this spatial smearing further reduces their contribution to any spurious signals. On all relevant metrics, the golden spiral design clearly outperforms our earlier aperiodic fiber where we obtained an experimental peak sidelobe level of -4.8 dB while maintaining a superior packing fraction ~($~\Lambda_{\mathrm{Fermat}}~= 11.8 ~\mu$m c.f. $\Lambda_{\mathrm{PR}} = 15 ~\mu$m). 

\begin{figure}[t]
  \centering
  \includegraphics[scale = 1]{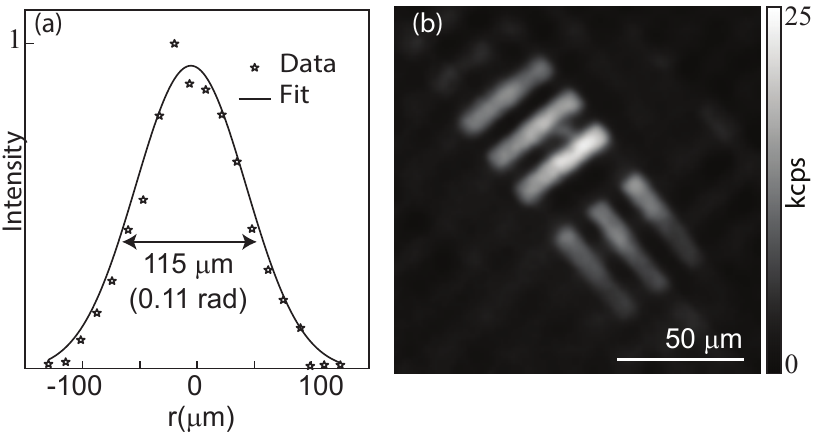}
  \caption{a) Measured relative intensity of the central lobe across the FoV. b) Two-photon image of an USAF target .
\label{fig:Images}}
\end{figure}

We further confirm that the transmission matrix of this fiber has no significant off-diagonal elements and use this property to apply simple linear and quadratic phase ramps on the SLM to translate the focal spot in the transverse and axial directions respectively. 
Figure~\ref{fig:Images}(a) depicts the intensity of the focal spots formed at various distances from the center of the FoV. This results in a Gaussian shaped intensity envelope which is consistent with the diffracted intensity pattern emerging from a single core \cite{Sivankutty2016OLprbundle}. To highlight the suitability of this fiber design for nonlinear imaging, we image an USAF test chart coated with fluorescein in the transmission scheme. The experimental setup now includes the two-photon detection module highlighted in the dashed lines in Figure~\ref{fig:Setup}(c). In order to prevent any bias due to the spatial selectivity of the generated signal, we image the entire backfocal plane of the collection objective (Olympus 20x, 0.48 NA) on to a single pixel detector. This corresponds to collecting any light generated from a large area ($1.21$ $\mathrm{mm^2}$) in the object plane. We use the SLM to perform a raster scan of the focal spot over a region of $\approx 145 \times 145 ~\mu$m wide, and integrate the generated two-photon signal on a PMT (H7421, Hammamatsu). The imaging rate is limited to 7 Hz by the update rate of the SLM. The scanning could as well be replaced by conventional galvanometric mirrors as in \citep{Andresen2013} utilizing the very large memory effect to overcome this speed limitation. The high contrast of the central pattern is clearly visible in Figure~\ref{fig:Images}(b). At the edges of the FoV, we do observe a very weak ghost image and this is consistent with the fact that the artifacts are at the limit of our estimated FoV. 

In our proof of concept implementation, the MCF does not have a sufficient collection efficiency in the epi-direction since there is no secondary cladding. We do not envision any conceptual difficulty in realizing such a structure with an additional fluorine-doped jacket. This eventually would enable high sensitivity endoscopic detection. The deterministic tiling of the cores would also enable the inclusion of stress rods to control the polarization state of the field \citep{Kim2015}. This is a major advantage compared to aperiodic MCFs made by rotating the fibers \cite{Sivankutty2016OLprbundle} or the groups of fibers \cite{kim2018semi} where realizing a common polarization axis for all the cores would be challenging. Furthermore, the optical transfer function (OTF) of the Fermat spiral-based design has a significantly smoother support (however not continuous). A well-behaved OTF would open avenues in computational methods such as phase retrieval and deconvolution. Such imaging properties of the Fermat spiral will examined more in detail in an upcoming publication.

In this report we have presented a novel MCF for the reduction of sidelobes, particularly for application in lensless two-photon imaging. We demonstrate atleast a 10.9 dB reduction in the sidelobes resulting in two-photon signal contrast (central and peak sidelobe) of $\geq$ 20 dB. In comparison, the corresponding  two-photon contrast factor for a periodic fiber would be 2.9 dB \citep{Andresen2013} and  9.6 dB in the case our earlier reported aperiodic fiber \cite{Sivankutty2016OLprbundle}. We expect that the highly contrasted PSF would also lend itself to numerical techniques to further enhance the quality of the images. Hence the Fermat spiral design combines the advantages of the hexagonal array such as high packing fraction, no inter-core coupling, infinite memory effect and low group delay dispersion with highly diminished sidelobes, a centrally peaked PSF and a FoV limited only by diffraction itself.

\paragraph{FUNDING}Agence Nationale de la Recherche (ANR-11-EQPX-0017, ANR-10-INSB-04-01, ANR 11-INSB-0006, ANR-14-CE17-0004-01), Aix-Marseille Universit\'e (ANR-11-IDEX-0001-02), Universit\'e Lille 1 (ANR-11-LABX-0007), Institut National de la Sant\'e et de la Recherche M\'edicale (PC201508), SATT Sud-Est (GDC Lensless endoscope), CNRS/Weizmann ImaginNano European Associated Laboratory, the Ministry of Higher Education and Research, Hauts de France council, European Regional Development Fund (CPER Photonics for Society P4S).

We thank Miguel A. Alonso for discussions, R\'emi Habert and Karen Delplace for technical assistance.

\end{document}